\documentclass[twocolumn,showpacs,preprintnumbers,amsmath,amssymb]{revtex4}
\usepackage{graphicx}
\usepackage{caption2}
\usepackage{dcolumn}
\usepackage{bm}
\newcommand\figcaption{\def\@captype{figure}\caption}

       \def\dt{\delta}
       \def\Dt{\Delta}
    
\def\ie{{\it i.e. }}   \def\cf{{\it cf. }}
  \def\etal{{\it et. al. }}
 \def\fc{\frac}
\def\bcc{\begin{center}} \def\ecc{\end{center}}
\def\beq{\begin{equation}} \def\eeq{\end{equation}}
\def\bea{\begin{eqnarray}} \def\eea{\end{eqnarray}}
\def\beqa{\begin{eqnarray}}  \def\eeqa{\end{eqnarray}}
 \def\la{\langle} \def\ra{\rangle}
\def\btbl{\begin{tabular}} \def\etbl{\end{tabular}}
\def\bpm{\begin{pmatrix}} \def\epm{\end{pmatrix}}
\def\btbb{\begin{tabbing}} \def\etbb{\end{tabbing}}
\def\btm{\begin{itemize}} \def\etm{\end{itemize}}
  \def\hs{\hskip}
  \def\nbr{\nonumber}
\def\f{\left}  \def\g{\right}  
\def\ched{\end{CJK*} \end{document}} 
     \def\bft{\overline{\ft}}

  \def\ft{t_{\rm fr}}
\def\ed{\end{document}}
\def\bpm{\begin{pmatrix}} \def\epm{\end{pmatrix}}
\def\ben{\begin{enumerate}} \def\een{\end{enumerate}}
\def\btb{\begin{tabular}} \def\etb{\end{tabular}}
\def\btbb{\begin{tabbing}} \def\etbb{\end{tabbing}}
 \def\EE{e$^+$e$^-$\hskip1pt }

\begin{document}

\title{The Narrowing of Charge Balance Function\\ and Hadronization Time
in Relativistic Heavy Ion Collisions}
\author{Du Jiaxin,\quad Li Na  \ and  \  Liu Lianshou\footnote{Email:
liuls@iopp.ccnu.edu.cn}} 
\affiliation{Institute of Particle Physics, Huazhong Normal
University, Wuhan 430079, China}

 \begin{abstract}
The widths of charge balance function in high energy hadron-hadron
and relativistic heavy ion collisions are studied using the Monte
Carlo generators PYTHIA and AMPT, respectively. The narrowing of
balance function as the increase of multiplicity is found in both
cases. The mean parton-freeze-out time of a heavy-ion-collision
event is used as the characteristic hadronization time of the
event. It turns out that for a fixed multiplicity interval the
width of balance function is consistent with being independent of
hadronization time.
\end{abstract}


\maketitle

\section{Introduction}
The relativistic heavy ion collision experiments at CERN-SPS and
especially at the relativistic heavy ion collider RHIC in
Brookhaven National Lab provide clear evidence for the production
of a dense matter in the collision processes \cite{evidence}. The
central question is whether this matter is purely hadronic or has
been going through a quark-parton phase. There exist experimental
evidences in favor of the existence of a quark-parton phase at the
early stage of collision processes \cite{qgp}, but in view of the
importance of this question, further confirmation is needed.

Recently, the  rapidity correlation between oppositely charged
particles, which has been used in \EE~\cite{ee} and hadron-hadron
collisions~\cite{hh} to study the hadronization in these
processes, is proposed~\cite{clock}  as a measure of the
hadronization time in relativistic heavy ion collisions. It is
argued that if the system produced in heavy ion collisions has
undergone a quark-parton phase, the hadronization will occur at a
later time and, therefore, the temperature will be lower and the
diffusive interaction with other particles will be lesser than
those in the direct hadronization without going through a
quark-parton phase. These will result in a narrower charge balance
function for a system with quark-parton phase than that without
such a phase.

Two heavy ion experiments~\cite{star, na49} have measured the
balance function at various centralities and for different
colliding nuclei. A narrowing of the balance function is indeed
observed with increasing centrality of the collision and with
increasing size of the colliding nuclei. These observations are
consistent with the assumption that the narrowing of balance
function is correlated with late hadronization.

On the other hand, recently it is reported~\cite{na22} that in
hadron-hadron collisions at $\sqrt s=22$ GeV the balance function
also becomes narrower as the increasing of multiplicity.
Therefore, whether the observed narrowing of balance function in
relativistic heavy ion collisions is due to late hadronization or
is simply due to the multiplicity effect is an open question.

In this letter this question is examined using the Monte Carlo
generators PYTHIA \cite{pythia} and AMPT \cite{ampt}. The former
is a standard Monte Carlo generator with string fragmentation as
hadronization scheme. There is not any quark-parton phase in this
model and the hadronization is almost instantaneous. On the other
hand, the latter is a ``multi-phase'' model, with a transport of
quark-parton before hadronization.

The results from PYTHIA will be presented in section II. The
hadronization time in AMPT model is discussed in section III, and
its connection with the width of balance function is presented in
section IV. Section V is summary and discussion.

\def\yw{Y_{\rm w}}   \def\dtyw{\la\dt y\ra_{\yw}}
\section{The width of balance function in PYTHIA model}

The balance function  is defined as~\cite{na49}
\begin{eqnarray}
 B(\dt y|Y_{\rm w}) &=& \frac{1}{2}\left[
 \frac{\la n_{+-}(\dt y)\ra-\la n_{++}(\dt y)\ra}{\la
 n_+\ra}\right.    \nonumber \\
 & &\hs15mm \left.+\frac{\la n_{-+}(\dt y)\ra-\la n_{--}(\dt y)\ra}{\la
n_-\ra}\right],\end{eqnarray} where, $n_{+-}(\dt y)$, $n_{-+}(\dt
y)$ and $n_{++}(\dt y)$, $n_{--}(\dt y)$ are the numbers of pairs
of opposite- and like-charged particles satisfying the criteria
that they fall into the rapidity window $Y_{\rm w}$ and that their
relative rapidity equals $\dt y$; $n_+$ and $n_-$ are the numbers
of positively and negatively charged particles, respectively, in
the interval $Y_{\rm w}$.

The balance function $B(\dt y|Y_{\rm w})$ represents the
probability that the balancing charges are separated by $\dt
y$~\cite{clock}. The mean of $\dt y$~\cite{na49}
 \beq \la\dt y\ra_{\yw}= \frac{\sum_i B( \delta
y_i|\yw)\delta y_i}{\sum_i B( \delta y_i|\yw)} \eeq is defined as
the {\it width of  balance function}.

\def\hfmb{\hfill\mbox {}}
Proton-proton collision events are generated at four c.m. energies
------ 22, 64, 130 and 200 GeV using PYTHIA5.720 generator. The event
number for each sample is 100,000. The widths $\la\dt
y\ra_{\infty}$ of balance function  in the full phase space are
calculated for different (charged) multiplicity bins and plotted
in Fig.~1.
\begin{figure}
\centering
\includegraphics[width=2.5in]{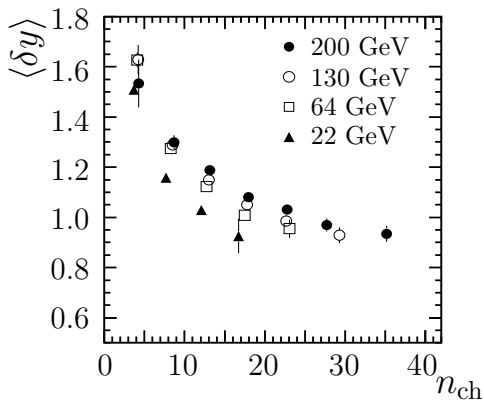}
\caption{\label{Fig. 1} The width of full-phase-space balance
function for different multiplicities in p-p collisions at
$\sqrt{s}=22,\; 64,\;130,\;200$~GeV.\hfill\mbox {}}
\end{figure}

It can be seen from the figure that in this  model even for p-p
collision, where no quark-parton phase is expected and the
hadronization is almost instantaneous, the width of  balance
function  decreases with the increase of multiplicity, \ie\ the
width of balance function is narrower for higher multiplicity.
This effect has nothing to do with hadronization time.

On the other hand, by definition balance function measures the
correlation length between oppositely charged particles. For
comparison we have calculated the standard 2-particle correlation
function~\cite{Foa} of oppositely charged particles 
\bea &&\hs-12mm R^{+-}(y_1,y_2) = \fc{1}{2} \f(
\fc{\rho^{(2)}(y_1^{+},y_2^{-})}{\rho^{(1)}(y_1^{+})\rho^{(1)}(y_2^{-})}\g.
\nbr\\ && \hs21mm +
\f.\fc{\rho^{(2)}(y_1^{-},y_2^{+})}{\rho^{(1)}(y_1^{-})\rho^{(1)}(y_2^{+})}
\g)-1\eea for different multiplicities in p-p collision at c.m.
energy $\sqrt s =200$ GeV, for $y_1=0$, $y_2=y$. The results
plotted in Fig.2 show that the width of $R$ is consistent with
being independent of multiplicity. A possible explanation of the
width of $R$ is cluster decay. Comparing with the definition of
balance function, Eq.\;(1), we see that it is the difference
between the correlations of opposite- and like-charged particles
that shows a clear multiplicity dependence, which is 
unrelated with cluster decay, and is mainly due to the string
fragmentation mechanism implemented in PYTHIA model.

\begin{figure}
\centering
\includegraphics[width=2.5in]{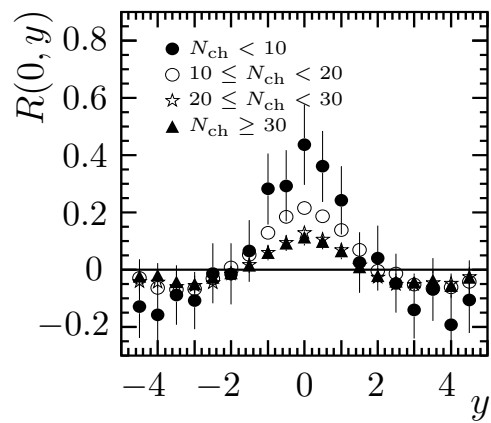}
\caption{\label{Fig. 2} The 2-particle correlation function
$R^{+-}(0,y)$ of oppositely charged particles as function of $y$
for different multiplicities in p-p collision at c.m. energy
$\sqrt s = 200$~GeV.\hfmb}
\end{figure}

It can also be seen from Fig.\;1 that the width of balance
function depends on collision energy. For the same multiplicity,
the higher the collision energy is, the wider the width of
balance function.

\begin{figure}
\centering
\includegraphics[width=2.5in]{figures/fig3.epsi}
\caption{\label{Fig. 3} The rapidity distribution of p-p collision
at $\sqrt{s}=22,\; 64,\;130,\;200$~GeV.\hfmb}
\centering
\includegraphics[width=2.5in]{figures/fig4.epsi}
\caption{\label{Fig. 4} The width of balance function in the
rapidity region $[-3,\, 3]$ for different multiplicities in p-p
collision at $\sqrt{s}=22,\; 64,\;130,\;200$~GeV.\hfmb}
\end{figure}

However, it should be noticed that the full rapidity region is
wider for higher energy, \cf Fig\;3. In order to get rid of the
influence of the  width of rapidity region we calculate the width
of balance function in the region $-3 \leq y\leq 3$ for all four
energies. The results, presented in Fig.~4, show that, when the
(average) rapidity density $\fc{\Dt n}{\Dt y}$ is the same, the
width of  balance function  is almost independent of energy,
especially for high $\fc{\Dt n}{\Dt y}$. That is, in hadron-hadron
collisions the width of  balance function  depends essentially
{\it only} on multiplicity and is consistent with being
independent on energy.

How does the width of  balance function  behave in nucleus-nucleus
collisions will be studied in the next two sections.

\section{Hadronization time in AMPT model}
The Monte Carlo generator AMPT1.11 is a multi-phase transport
model, which contains a quark-parton transport phase before
hadronization. The initial spatial and momentum distributions of
hard partons and soft string excitations are obtained from the
HIJING~\cite{hijing} model. The parton cascade follows Zhang's
parton-cascade (ZPC) model~\cite{zpc}, which is based on two-body
pQCD scattering with screening masses. When interaction ceases,
the partons are recombined with their parent strings to form
hadrons according to LUND string fragmentation
mechanism~\cite{pythia}.        
Then the scatterings among the resulting hadrons are described by
a relativistic transport (ART) model~\cite{art} which includes
baryon-baryon, baryon-meson and meson-meson elastic and inelastic
scatterings.

In our calculation the parton cross section is chosen to be 10 mb.
If the colliding nuclei are large and the energy is high, then the
parton cascade will last long enough to make the parton
distribution arrive at an equilibrium distribution~\cite{ko}.
However, no equilibrium thermodynamics has been included in the
model. In particular, there is not any equilibrium phase
transition from parton phase to hadron phase. Therefore, there is
no unique hadronisation time for the whole system. Each parton has
its own hadronisation time, or freeze out time $\ft$.

\def\npt{n_{\rm parton}}
In order to study the correlation, if any, between the width of
balance function and the hadronization time, we use the event mean
of parton freeze out time $\ft$ \beq \bft = \fc {1}{\npt}
\sum_{i=1}^{\npt} {\ft}_i \eeq as the characteristic hadronization
time for an event, where $\npt$ is the number of partons in the
event, ${\ft}_i$ is the freeze out time of the $i$th parton.

\begin{figure}
\centering
\includegraphics[width=3.2in]{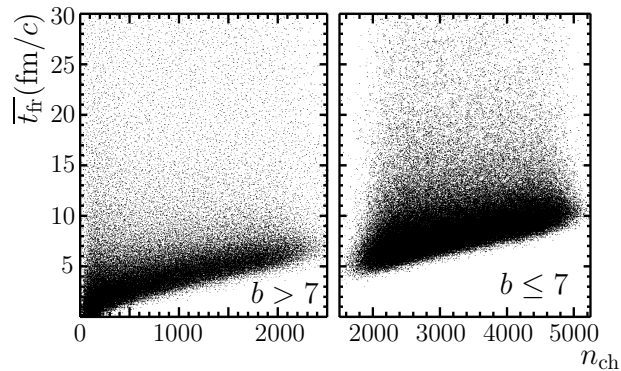}
\caption{\label{Fig. 5} Scattering plots of $\bft$ vs. $n_{\rm
ch}$ in Au-Au collision at $\sqrt{s_{NN}}=200$~GeV for two
different centralities.\hfmb}
\end{figure}

The AMPT1.11 (default) generator is utilized to generate  Au-Au
collision events at $\sqrt{s_{NN}}=200$\;GeV. Two event samples
with 250,000 and 770,000 events, respectively, are generated for
two centralities $b\leq 7$\;fm and $b>7$\;fm.

In Fig's.\;5 are shown the scattering plots of \,$\bft$\, vs.
$n_{\rm ch}$ in Au-Au collisions at $\sqrt{s_{NN}}=200$~GeV for
the two different centralities ------ $b\leq 7$\;fm and $b>7$\;fm,
respectively. It can be seen that in central collisions ($b\leq
7$\;fm) $\bft$ is larger than 5\;fm, while in peripheral
collisions ($b > 7$\;fm) $\bft$ is concentrated at $\bft<5$\;fm.
That is, central collision events hadronize later than peripheral
ones.

The distributions of event-mean parton freeze out time $\bft$ for
the two different centralities in Au-Au collision at
$\sqrt{s_{NN}}=200$~GeV are shown in Fig.\;6.

\begin{figure}   
\includegraphics[width=2.5in]{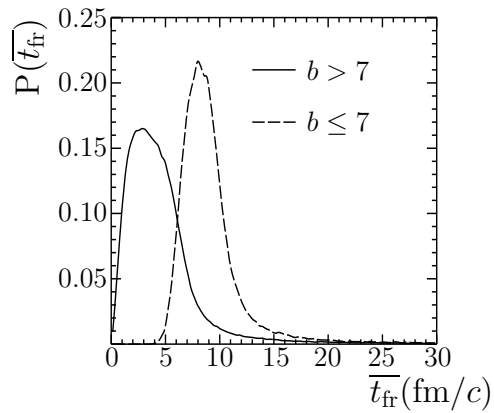}
\caption{\label{Fig. 6} The distributions of event-mean parton
freeze out time $\bft$ for two different centralities in Au-Au
collision at $\sqrt{s_{NN}}=200$~GeV.\hfmb}
\end{figure}

\section{Relation between the width of Balance function and hadronization time}

In order to study the correlation between the width of balance
function and the characteristics of single event --- event-mean
parton freeze out time $\bft$ and/or multiplicity $n_{\rm ch}$,
each centrality sample is divided into sub-samples according to
the intervals of mean parton freeze-out time $\bft$ and the
resulting sub-samples are further divided into sub-samples by
different multiplicity intervals.

The widths of  balance function in the rapidity region $Y_{\rm
W}=[-3,3]$ for different intervals of mean parton freeze-out time
$\bft$ and multiplicity $n_{\rm ch}$ are shown in Fig.\;7. The
abscissa of the figure is the average multiplicity in the
corresponding multiplicity interval.

It can be seen from the figure that in all cases the width of
balance function decreases with the increasing of multiplicity,
while in the same multiplicity interval, the width of balance
function  is consistent of being constant, independent of the
hadronization time.  Therefore, to use the narrowing of balance
function in relativistic heavy ion collision as a measure of
hadronization time and as a signal of QGP is doubtful.

\begin{figure}
\centering
\includegraphics[width=3.0in]{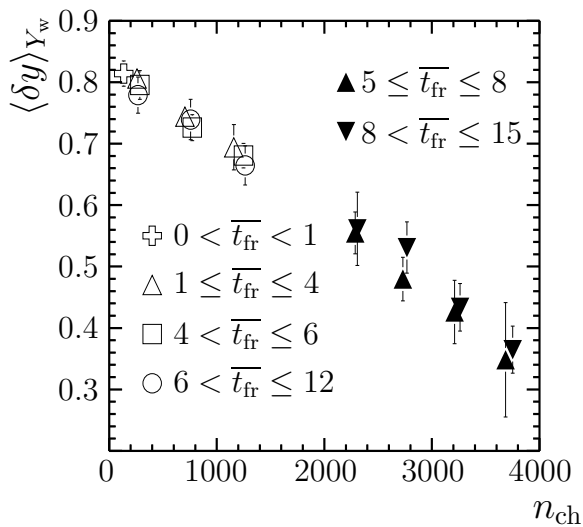}
\caption{\label{Fig. 7} The widths of  balance function  in the
rapidity region $Y_{\rm W}=[-3,3]$ for different mean parton
freeze-out time $\bft$ intervals versus multiplicity $n_{\rm ch}$
in Au-Au collision at $\sqrt{s_{NN}}=200$~GeV.\hfmb}
\end{figure}

\section{Summary and discussion}

It is found using PYTHIA Monte Carlo that the width of charge
balance function decreases with the increasing of multiplicity in
high energy hadron-hadron collisions, where the hadronization is
almost instantaneous.

The relation between the hadronization time and the width of
charge balance function in relativistic heavy ion collisions is
examined using the default AMPT1.11 Monte Carlo generator.  The
mean parton freeze out time of an event is used as the
characteristic hadronization time of the event. The narrowing of
balance function as the increase of multiplicity is found to exist
also for relativistic heavy ion collisions, while for a fixed
multiplicity interval the width of balance function is consistent
with being independent of hadronization time. Therefore, to use
the narrowing of balance function in relativistic heavy ion
collisions as a measure of hadronization time and as a signal of
QGP is doubtful.

It should be noticed that AMPT model is a multi-phase {\it
transport}\, model. There is no equilibrium thermodynamics
included in the model. In particular, there is no equilibrium
phase transition, and consequently no unique hadronization time
for an event. To use the average of parton freeze out time in an
event as the characteristic hadronization time of the event is a
crude approximation. Therefore, the observation made in the
present work on the independence of the width of balance function
on hadronization time is only a first step. Further investigation
along this line is needed.

{\bf Acknowledgement} \ This work is supported by NSFC under
project 10375025 and by the Cultivation Fund of the Key Scientific
and Technical Innovation Project of Ministry of Education of China
NO CFKSTIP-704035.

\end{document}